\documentclass[sigconf]{acmart}
\usepackage[textsize=scriptsize,disable=True]{todonotes}
\usepackage{mathtools, nccmath}

\usepackage{xpatch}
\xpatchcmd{\NCC@ignorepar}{%
\abovedisplayskip\abovedisplayshortskip}
{%
\abovedisplayskip\abovedisplayshortskip%
\belowdisplayskip\belowdisplayshortskip}
{}{}
\renewcommand\footnotetextcopyrightpermission[1]{}
\usepackage[nodayofweek,level]{datetime}
\RequirePackage{soul}
\usepackage{multirow}
\usepackage{fancybox}
\RequirePackage{url}
\RequirePackage[inline]{enumitem} \setlist{nosep}
\dbltextfloatsep \textfloatsep
\intextsep \textfloatsep
\setcopyright{acmcopyright}
\copyrightyear{2022}
\acmYear{2022}
\setcopyright{acmcopyright}\acmConference[WSDM '22]{Proceedings of the Fifteenth ACM International Conference on Web Search and Data Mining}{February 21--25, 2022}{Tempe, AZ, USA}
\acmBooktitle{Proceedings of the Fifteenth ACM International Conference on Web Search and Data Mining (WSDM '22), February 21--25, 2022, Tempe, AZ, USA}
\acmPrice{15.00}
\acmDOI{10.1145/3488560.3498511}
\acmISBN{978-1-4503-9132-0/22/02}
\renewcommand\footnotetextcopyrightpermission[1]{}

\newcommand{\name}{\texttt{SANDS}}
\newcommand{\usadata}{\texttt{StanceUS}}
\newcommand{\indiadata}{\texttt{StanceIN}}
\newcommand{\fullname}{{\bf S}tance {\bf A}nalysis
via {\bf N}etwork {\bf D}istant {\bf S}upervision}

\begin{document}



\def\ztitle{Semi-supervised Stance Detection of Tweets\\ Via Distant Network Supervision}
\title{\ztitle}
\author{$^1$Subhabrata Dutta, $^2$Samiya Caur, $^3$Soumen Chakrabarti, $^2$Tanmoy Chakraborty}
\affiliation{%
   \institution{$^1$Jadavpur University, India; $^2$ IIIT-Delhi, India; $^3$ IIT Bombay, India}
}

\begin{abstract}
Detecting and labeling stance in social media text is strongly motivated by hate speech detection, poll prediction, engagement forecasting, and concerted propaganda detection.  
Today's best neural stance detectors need large volumes of training data, which is difficult to curate given the fast-changing landscape of social media text and issues on which users opine. Homophily properties over the social network provide strong signal of coarse-grained user-level stance. But semi-supervised approaches for tweet-level stance detection fail to properly leverage homophily. In light of this,
We present \name, a new semi-supervised stance detector.
\name{} starts from very few labeled tweets.
It builds {\em multiple deep feature views} of tweets.
It also uses a {\em distant supervision signal} from the social network to provide a surrogate loss signal to the component learners.
We prepare two new tweet datasets comprising over 236,000 politically tinted tweets from two demographics (US and India) posted by over 87,000 users, their follower-followee graph, and over 8,000 tweets annotated by linguists.
\name\ achieves a macro-F1 score of $0.55$ ($0.49$) on US (India)-based datasets, outperforming 17 baselines (including variants of \name) substantially, particularly for minority stance labels and noisy text. Numerous ablation experiments on \name\ disentangle the dynamics of textual and network-propagated stance signals. 
\end{abstract}

\maketitle

\section{Introduction}
\label{sec:intro}

Social media is regarded as a barometer of modern society's emotional state.  Billions of social media users express their stances toward events, social issues, and political parties in their tweets, Facebook articles, or blogs.  The `target entity' may not be explicitly mentioned in the text expressing the stance.  
Automatic stance detection is a strongly motivated mining operation on social media and networks 
\citep{MOHAMMAD16.232}.  Some applications include hate speech detection \citep{DBLP:conf/wassa/GrimmingerK21}, poll prediction \citep{grvcar2017stance}, and rumor veracity detection \citep{dungs-etal-2018-rumour}. The importance of political stance analysis over Twitter-like platforms has increased dramatically in recent times owing to several phenomena --- sharp increase in partisanship among users, malicious efforts of organized groups to distort popular opinion at a large-scale, etc.

\begin{figure*}
\centering
\includegraphics[width=\textwidth]{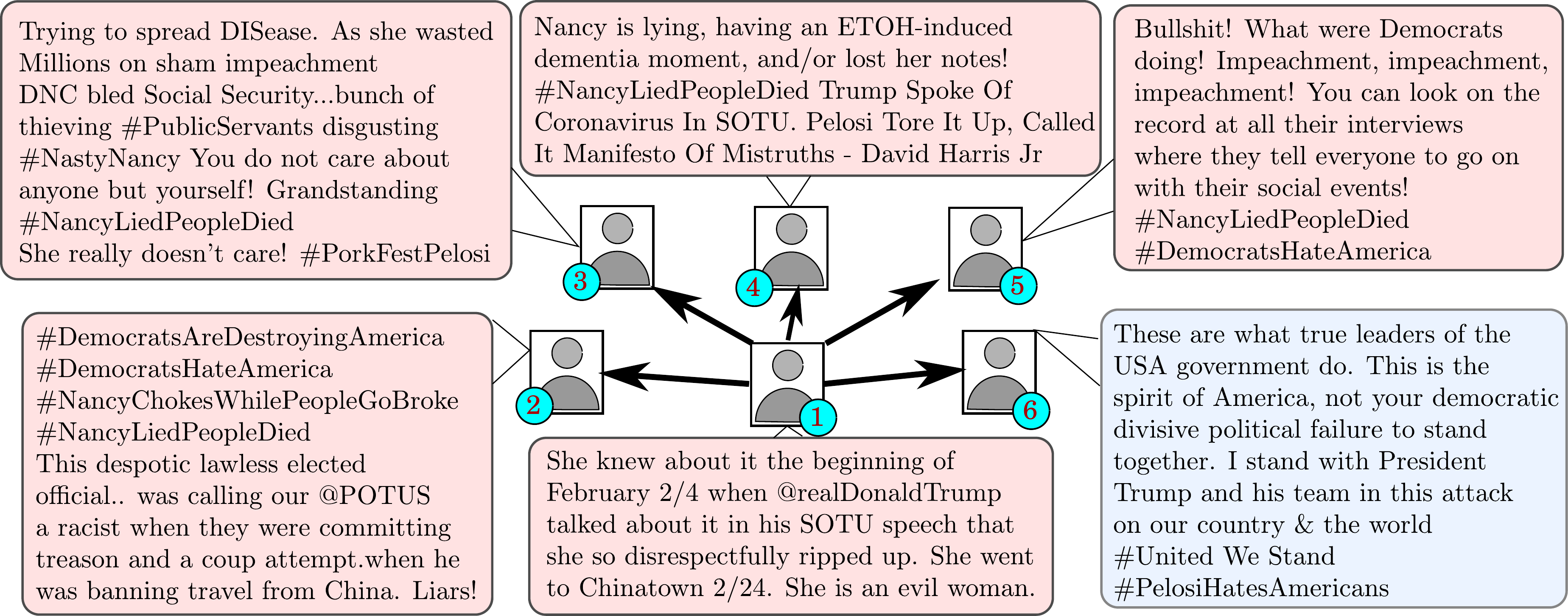}
\caption{Stance homophily in Twitter. User $\mathbf{1}$ follows users $\mathbf{2}$, $\mathbf{3}$, $\mathbf{4}$, $\mathbf{5}$, and $\mathbf{6}$. All these users carry similar opinion, with user $\mathbf{6}$'s tweet showing support to the Republicans while the rest are anti-Democrat. While the tweet posted by user $\mathbf{1}$ does not link any entity related to Republican or Democrats directly to some polarity words (thereby making the stance classification difficult), a classification framework with the knowledge of the rest of the tweets can break the ambiguity.}
\label{fig:SampleTweets}
\end{figure*}

\textbf{Leveraging homophily in stance detection.} A large number of approaches have been proposed for text-based stance detection \cite{MOHAMMAD16.232,BICE,CrossNet}.  Relatively few approaches recognize and exploit the fact that text in social networks is accompanied by rich graph-structured metadata, e.g., friends, followers/followees, retweets and replies, hashtags, text-author associations, etc.~\citep{network-for-stance-2016,unsup-user-stance_2020}. It is well-known that {\em homophily} (friends have similar taste) and {\em social balance} (enemy of an enemy is a friend, etc.) are pervasive in social networks.  Therefore, these signals have the potential to improve the accuracy of stance prediction.
Consider the tweets shown in Figure~\ref{fig:SampleTweets}.  The stance labels in this example include pro-Republican, anti-Republican, pro-Democrat, anti-Democrat, and neutral.  Tweets from users 2--5 clearly express an anti-Democrat stance, while user 6's tweet is pro-Republican.  It may be non-trivial to correctly label the tweet of user 1 (because `she' is not identified, and the only handle @realDonaldTrump is uninformative) unless we see whom user 1 follows and what tweets they write. 
However, previous works connecting network dynamics for stance classification mostly deal with user-level stance analysis. An overall stance of the user often does not reflect in the tweet. For example, a Democrat supporter can tweet something pro-democrat or anti-republican. 
Volatile users may often switch their stances depending on the issue. For example, among the tweets we collected and annotated for the analysis presented in this work, about $8\%$ show stance switch. On the other hand, supplementing local, tweet-level features with homophily-driven features might add bias towards the majority stance of the neighbor nodes. Instead, we seek to use homophily as a navigator of distant supervision. The end results are simple, tweet-level stance classifiers that rely on network-level information only at training time.


\textbf{Scarcity of labeled data for stance classification.} A key hurdle in our setting is the paucity of human-labeled data. Neural text processors are among the best for stance detection; however, they need large volumes of training data. The rapid pace of information generation and consumption over social media leads to the emergence of completely new entities and concepts (persons, events, issues, etc.),  too fast for the curation of human-labeled high-quality data for each scenario.  Semi-supervised and active learning are common coping mechanisms. Starting from a small set of instances manually labeled with ground-truth (`gold instances'), they expose the learner to progressively larger sets of instances that are automatically and manually labeled, respectively.  Label sampling decisions are not usually informed by an overlying network.

\textbf{Proposed method: \name} Our central research question is the manner in which network signals like homophily can improve semi-supervised stance detection.  Our investigation results in a system, called \name{} (\fullname) that starts from a few high-quality seed instances, obtains noisy labeling guidance from homophily (between a user and her followers), uses multiple views of tweet content with customized feature extraction models, and iteratively train the component learners.  

In more detail, \name{} uses three kinds of textual feature extractors, designed to fit their semantics in the context of 
Twitter ---  (i)~A simple symmetric set aggregated encoding is used for hashtags. (ii)~Short-range contextual text representation is captured through a convolutional network. (iii)~Longer range textual context is captured through a 
bi-LSTM.  These are combined into two main learners/predictors with somewhat different capabilities and strengths.  Rather than throwing these into a standard co-training or disagreement-based learning scheme, we also make use of network information along with a homophily assumption that {\em followees of a user tend to have the same stance as the user}.  Therefore, the predictors are also applied to recent tweets by followees of the user who posted the given tweet, and their majority vote turned into a suitable loss for the other learner.

This novel combination of {\em network-driven distant supervision} and {\em synthetic feature view separation} makes \name{}'s predictive accuracy considerably superior to many recent and competitive baselines as well as ablations.  \name{} is particularly effective at improving the accuracy of minority class labels.

\paragraph{Two new datasets and the superiority of \name}  As part of this work, we offer two large collections of politically tinted tweets from the two most vociferous social media with political content: $\mathbf{59,684}$ and $\mathbf{176,619}$ tweets from US and India-based users, respectively, along with the corresponding inter-user follow relations; $3,822$ and $4,185$ tweets respectively among these collections are manually annotated with corresponding stance labels. While using only $1,500$ samples as labeled data, \name{} achieves $\mathbf{0.55}$ and $\mathbf{0.47}$ macro-F1 scores on US and India-based datasets, surpassing the best baseline by (absolute) \todo{CHECK done} 4\% and 5\%, respectively. 

\paragraph{Summary of our major contributions:}
\begin{itemize}[leftmargin=*]
\item We propose a novel stance classification framework for tweets, \name{}, which employs distant supervision using follow network information to learn tweet stances with frugal labeled data.
\item Using separate neural feature extractors to focus on local and global contextual features from tweet texts separately, we formulate a learning strategy facilitated by distant network supervision that iteratively trains multiple models with \todo{unexplained jargon? rephrased} two-phased view separations: local vs. global textual features within the tweet and candidate tweet vs. followee tweets over the network.
\item We present two large collections of tweets from US and India-based users, along with manually annotated labels for political stance on subsets of each.
\item We perform extensive experiments with multiple supervised and semi-supervised methods along with ablation variants to analyze signal importance.
\end{itemize}

\paragraph{Reproducibility:}
We detail the pre-processing and parameters of \name\ necessary to reproduce the results in the supplementary material. Also, the source codes are available on this anonymous repository: \url{https://github.com/LCS2-IIITD/SANDS}.


\section{Related work}
\label{sec:related_work}

{\bfseries Stance and target-dependent sentiment classification.}
Over\-lapping communities work on stance detection and target-dependent sentiment classification.  Given a target entity, the task of predicting the sentiment (positive, negative, or neutral) expressed over a tweet or any other form of text is well explored, since the pioneering work by \citet{td-sentiment-2011-acl}. \citet{multiTDSenti-twitter-2017-EACL} approached the problem of identifying sentiment toward multiple entities present in a tweet. Recent works have explored a variety of machine learning frameworks for the task: LSTM with attention mechanism \citep{tdsenti-lstm-att}, multi-task learning \citep{mttdsc}, dual transformer with graph learning \citep{tdsenti-2020-acl}, \emph{inter alia}. 


While the detection of stance (pro-, anti-, or none) towards a target expressed in a single tweet \citep{MOHAMMAD16.232} is  similar to target-dependent sentiment classification, the general literature of stance classification covers a much broader ground --- stance towards a rumor \citep{rumor-stance_2016, rumor-stance-2019, 8955493}, user stance prediction \citep{unsup-user-stance_2020, semisup-user-stance-2020-acl}, etc. \citet{MOHAMMAD16.232} presented a benchmark dataset for tweet stance classification task, focused on political tweets about the 2016 US Presidential Election. To explore stance detection in a multi-lingual scenario, \citet{multilingual-stance-dataset-2020} annotated Catalan and Spanish tweets regarding the Catalan referendum. \citet{conforti-etal-2020-will} presented a large-scale dataset of tweets annotated with stance expressed towards several corporate events in the pharmaceutical industry. 
\citet{MOHAMMAD16.232} showed that a linear SVM model with n-gram and several sentiment-based features achieve a consistent baseline for this task. Other approaches seek to develop sophisticated target-augmented learning of tweet representation using different neural network-based methods~\citep{BICE,CrossNet,TAN}.
Most of these datasets (and consequently, the models) follow a similar stance labeling strategy --- given a tweet and a specific target, the classes are {\em pro}-target, {\em anti}-target, or {\em none}. However, in the interest of higher-quality human annotation, we use a slightly different classification scheme in this work. For a given set of targets, we merge all the pro- and anti- labels in a single label set, and choose the most expressed opinion as the stance label. For example, given the target set as {\em Democrat} and {\em Republican}, our label set becomes pro-Democrat, anti-Democrat, pro-Republican, anti-Republican, and other/none; the tweet from user $6$ shown in Figure~\ref{fig:SampleTweets} is associated with a pro-Republican label following its major bias, though an anti-Democrat tone is also present.

 
{\bfseries User stance and network dynamics}
Some studies explored the problem of identifying the stance held by a user in general, instead of specific tweets. While tweets posted by a user signal the overall user stance, there are other possible metadata as well --- profile information, follow information, polls, etc. \citet{unsup-user-stance_2020} implemented a user-clustering approach to label them according to their stances. \citet{semisup-user-stance-2020-acl} explored a semi-supervised approach based on label propagation. \citet{network-for-stance-2016} tackled the problem of predicting future user stance based on user dynamics and network activity, in the aftermath of a major event. Their findings suggest that network-based features provide strong signals for predicting user stance. 

While the aforementioned studies mostly focused on stance at a micro level, the large-scale dynamics of user stance at the macro level is another focus of interest. \citet{Stance-homophily-Twitter} explored the homophily properties of the social network of Twitter based on opinion bias. For half a decade now, the increasing trend of ``echo chamber'' formation among Twitter and other social media users have led to users increasingly getting ``locked up'' within communities of similar opinion~\citep{echo-chamber-2018,echo-chamber-2}.
\citet{Stance-homophily-political-2019} observed inverse homophily, where some social ties are defined as ``reply-to-messages'' relationships. Conflicts of opinion among clustered user groups have been observed in platforms other than Twitter as well~\citep{10.1145/3178876.3186141, conflict-cluster}.

{\bfseries Co-training for text classification}
Co-training is a  semi-sup\-er\-vised learning approach where multiple classifiers `teach' each other based on learning from independent views of the data~\citep{cotrain/BlumM98, co-training-email-classify}. In the absence of naturally multi-view data, \citet{single-view-cotraining} proposed feature space partitioning to enforce different classifiers receiving a synthetic view-difference. \citet{bilingual-cotraining} exploited languages as views to apply co-training for sentiment classification of bilingual texts. \citet{cotraining-sentiment-mooc} applied co-training to classify sentiments over MOOC forum posts using a synthetic difference of views generated by classifier design. They used character-level and word-level processing of text to enforce view difference.

Our proposed method, \name, exploits view difference at both micro (tweet instance) as well as macro (network dynamics) levels. We extract feature signals within local and global contexts of the tweet separately using two independent classifiers, imposing a feature-level synthetic view-separation. On top of this, we use the homophily property in the user-user follow network as a distant supervision strategy, to present to the learners a parallel view of the tweet stance that is distinct from the text contents. Altogether, the classifiers leverage massive unlabelled data to minimize disagreement from these two levels of view separations, along with a small set of labeled data to minimize the task specific loss.

\section{Proposed Method}
\label{sec:method}

\subsection{Preliminaries and overview}
\label{subsec:prelims}

Let $\mathcal{U}$ be a set of users in a (partial) social network of Twitter, where $u_i\rightarrow u_j$ signifies user $u_i$ {\em follows} user $u_j$, $u_i, u_j \in \mathcal{U}$. The {followee} set of $u_i$ is defined as $f(u_i):=\{u_j|u_i\rightarrow u_j\}$. Let $\mathcal{D} := \{\tau^i_t|0\leq t \leq T\}$ denote the temporally ordered set of tweets over a period of time $[0, T]$, where $\tau^i_t$ is the tweet posted by user $u_i$ at time $t$.  Given a tweet $\tau^i_t$, we define its stance to be a categorical distribution over a specified label set~$\mathcal{L}$.
A small subset of $\mathcal{D}$ is manually annotated, denoted by $\mathcal{D}_s = \{(\tau^i_t, l^i_t)| \tau^i_t\in \mathcal{D}, l^i_t\in \{0,1\}^{|\mathcal{L}|} \}$, where $l^i_t$ represents the one-hot class label for tweet $\tau^i_t$. For example, the tweet from user~1 in Figure~\ref{fig:SampleTweets} has the corresponding stance label {\bf Anti-Democrat}, where the label set $\mathcal{L}$ is [{\bf Pro-Democrat}, {\bf Anti-Democrat}, {\bf Pro-Republican}, {\bf Anti-Republican}, {\bf Other}]. A stance classifier in our setting can be defined as a mapping $\mathcal{C}: y^i = \mathcal{C}(\tau^i_t|\theta)$, where $y\in (0,1)^{|\mathcal{L}|}$ is the predicted distribution of label probability, and $\theta$ is the set of parameters of the model.



The classical co-training paradigm requires the existence of two mutually independent views of the data \citep{cotrain/BlumM98}. In our case, the data source (a single tweet) does not provide any such natural partitioning. We seek to mitigate this by introducing two major design decisions. First, we design two separate classifiers to focus on local and global textual contexts of the tweet, thereby aggregating separate feature signals for stance polarity from a single tweet independently. Second, we exploit stance homophily over Twitter \citep{Stance-homophily-Twitter} to provide another avenue of incorporating multiple views; we assume that the stance polarity expressed by a user $u_i$ in tweet $\tau^i_t$ can be independently inferred from the polarities of the previous tweets from its followees,  $\{\tau^{f(u_i)_j}_{t-}\}$.

\subsection{Classification models}
\label{subsec:models}

Here, we describe the architectures of two classifiers we use for our task. Typically, a tweet consists of a piece of text body and one or more hashtags. For both our classifiers, these two input components are processed independently at the initial stages and fused before the final prediction. We represent the textual body of the tweet as a sequence of words $W_1, W_2, \cdots, W_n$ and the hashtags as a list $\#H_1, \#H_2, \cdots, \#H_m$. To learn the feature representations from the hashtags, both the classifiers use a similar strategy based on self-attention. However, to enforce learning pseudo-separate views from the text body, the classifiers use different methods of learning text representations: $\mathcal{C}_1$ uses a {\em convolutional} architecture focusing on short, contiguous segments of the text (i.e., local context) while $\mathcal{C}_2$ uses a {\em bidirectional LSTM-based} one which encodes distantly separated textual signals into one single representation (i.e., global context)\footnote{Experiments with pretrained BERT did not yield satisfactory results; see Section~\ref{subsec:classifier_pair}}.

\subsubsection{Learning hashtag representations} 
\label{subsubsec:hashtag}

We use an embedding layer to transform the list of one-hot representations of hashtags ${\#H_i}$ to a list of $d_H$ dimensional dense vectors $\mathbf{X^H}=\{x^H_i\}$. 

Next, to control the contribution of each of these vectors for the prediction task, we compute {\em scaled dot product attention} between them. For this, we compute the {\em query}, {\em key}, and {\em value} vectors from each of $x^H_i$ as, $q_i = \operatorname{ReLU}(\mathbf{W}_q x^H_i + \mathbf{B}_q);\ 
        k_i = \operatorname{ReLU}(\mathbf{W}_k x^H_i + \mathbf{B}_k);\ 
        v_i = \operatorname{ReLU}(\mathbf{W}_v x^H_i + \mathbf{B}_v)$,
where $\mathbf{W}_q, \mathbf{W}_k, \mathbf{W}_v$, $\mathbf{B}_q, \mathbf{B}_k, \mathbf{B}_v$  are learnable weight and bias matrices. We compute the attention weights and scale the representations for each hashtag as follows:
\begin{equation*}
\small
\begin{split}
    \alpha_{ij} &= \frac{\exp(q_i^{\top} k_j)}{\sum_j \exp(q_i^{\top} k_j)}; \qquad
    \hat{x}^H_i = \sum_j \alpha_{ij} v_j
\end{split}
\end{equation*}
Finally, we apply max-pooling over $\{\hat{x}^H_i\}$ to compute the combined representation of the hashtags, $\mathbf{Z}^H$.

\subsubsection{Convolutional feature extraction}
\label{subsubsec:conv_model}

Given a sequence of one-hot word representations $\{W_i\}$, an embedding layer maps them to $d_W$-dimensional vectors $\mathbf{X}^W=\{x^W_i\}$. Then, we apply three branches of consecutive 1d-convolution and max-pooling operations with window sizes $1$, $3$, and $5$ on $\mathbf{X}^W$. A single pair of convolution-pooling operation can be represented as
$\mathbf{h}_{i+1} = \operatorname{Maxpool}(\operatorname{ReLU}(\operatorname{Conv}(\mathbf{h}_i)))$.
For three successive such operations within a branch of window size $k$, $\mathbf{h}_0=\mathbf{X}^W$ and $\mathbf{h}_3=\hat{\mathbf{Z}}^W_{\text{conv}-k}$. Finally, outputs of each of these branches along with the hashtag representation $\mathbf{Z}^H$ are concatenated, normalized, and forwarded to the prediction feed-forward layer with softmax activation:\useshortskip
\begin{equation*}
\small
\begin{split}
    \mathbf{Z}_1 &= \operatorname{LayerNorm}([\mathbf{Z}^H:\mathbf{Z}^W_{\text{conv}-1}:\mathbf{Z}^W_{\text{conv}-3}:\mathbf{Z}^W_{\text{conv}-5}])\\
    y_1 &= \operatorname{Softmax}(\mathbf{W}_{p1} \mathbf{Z}_1 + \mathbf{B}_{p1})
\end{split}
\end{equation*}
where $\mathbf{W}_{p1}$ and $\mathbf{B}_{p1}$ are learnable kernel and bias matrices of the feed-forward layer, $(\cdot : \cdot)$ denote the concatenation operation, and $y_1\in (0,1)^{|\mathcal{L}|}$ is the predicted class probability. Figure~\ref{fig:my_label} in the supplementary material contains the detailed information flow.
\begin{figure}
    \centering
    \includegraphics[width=\columnwidth]{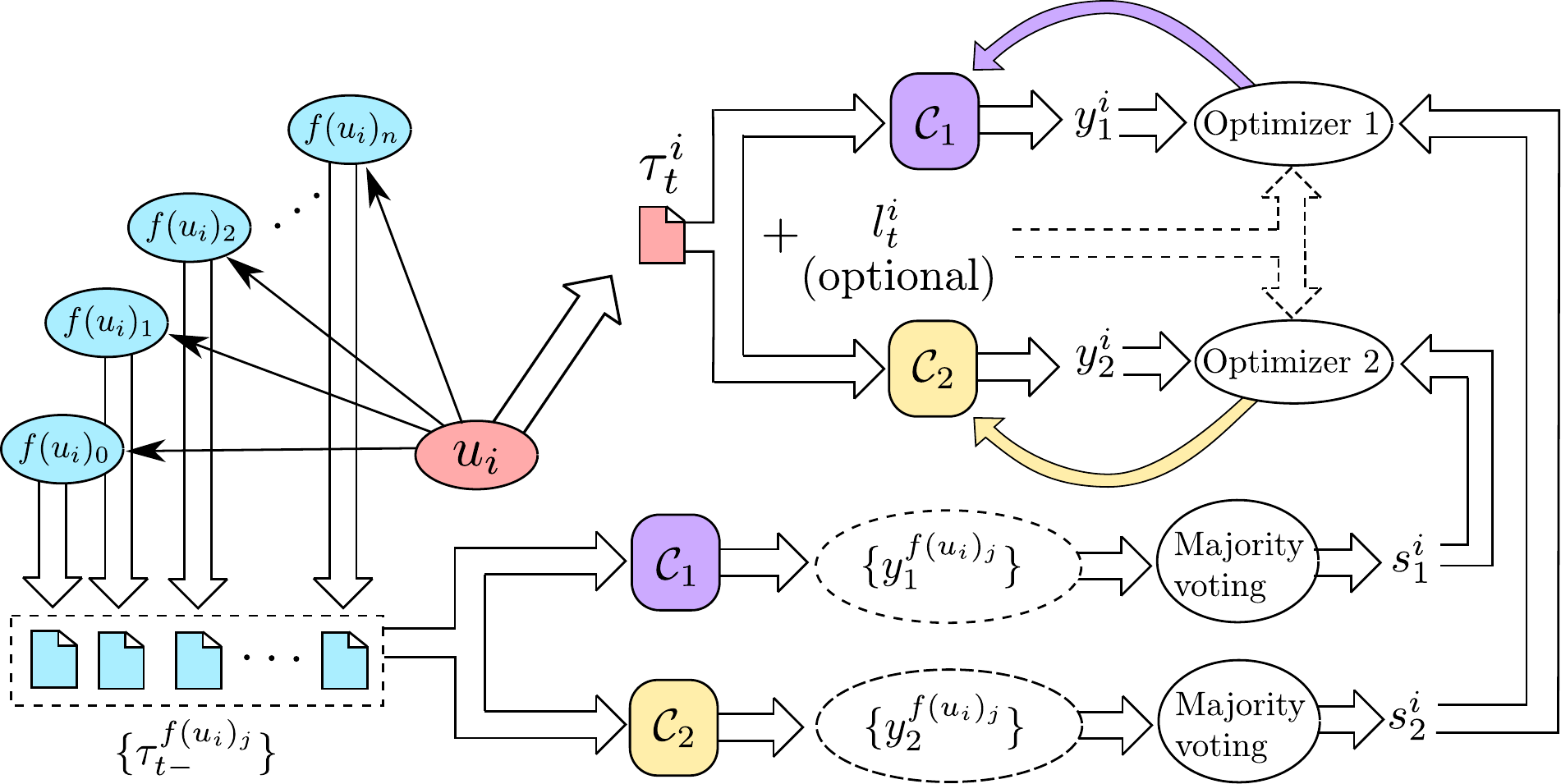}
    \caption{Training process of \name. Classifiers $\mathcal{C}_1$ and $\mathcal{C}_2$ independently predict the stance for a tweet from user $u_i$. In parallel, the classifiers generate pseudo-label sets $\{y^{f(u_i)_j}_1\}$ and $\{y^{f(u_i)_j}_1\}$, respectively from the recent tweets by followees of $u_i$. After majority voting, lables $s^i_1$ and $s^i_2$ are selected from pseudo-labels. $\mathcal{C}_1$ is optimized using the label generated by $\mathcal{C}_2$, and vice versa. Additionally, annotated labels $l^i_t$ (if present) of the tweet are also used to compute loss and optimize the classifiers.}
    \label{fig:training}
\end{figure}

\subsubsection{Bi-LSTM-based feature extraction}
\label{subsubsec:blstm_model}

Similar to its convolutional counterpart, the bidirectional LSTM-based module also uses an embedding layer to map one-hot words into fixed dimensional vectors of size $d_W$. 
A bidirectional LSTM layer then maps these vectors into an intermediate sequential representation $\mathbf{h}_{\text{blstm}}=\{h_i\}$, which, upon max-pooling along the sequence axis, produces the representation of the tweet body $\mathbf{Z}_{\text{blstm}}$. This is then concatenated with the hashtag representation, normalized, and passed to the prediction layer, similar to Section~\ref{subsubsec:conv_model}:
\begin{equation*}
\small
\begin{split}
    \mathbf{Z}_2 &= \operatorname{LayerNorm}([\mathbf{Z}^H:\mathbf{Z}_{\text{blstm}}])\\
    y_2 &= \operatorname{Softmax}(\mathbf{W}_{p2} \mathbf{Z}_2 + \mathbf{B}_{p2})
\end{split}
\end{equation*}

Henceforth, we denote the convolutional and bi-LSTM based models as $\mathcal{C}_1$ and $\mathcal{C}_2$, respectively. Additionally, each sub-block of operations in each model 
except the final feed-forward layer is followed by a dropout layer with probability $p_{\text{dropout}}\in (0,1)$ to handle overfitting and noise.

\subsection{Training method}
\label{subsec:training}
Our proposed training methodology in \name\ consists of two repeating phases. First, each of the classifiers $\mathcal{C}_1$ and $\mathcal{C}_2$ defined in Section \ref{subsec:models} is optimized separately on the labeled dataset $\mathcal{D}_s$ in a purely supervised manner. Then, we incorporate the full dataset (labeled as well as unlabeled) to jointly train the classifiers. In this phase, relying on the homophily of user opinions, the classifiers generate labeling of tweets depending on the network level view (i.e., recent tweets from followees) for each other. This way, the local context aggregating classifier $\mathcal{C}_1$ receives the alternate, global context-based, network-level view from $\mathcal{C}_2$, and vice versa. 

\subsubsection{Supervised phase}
\label{subsubsec:supervised_pretrain}

Given the annotated dataset $\mathcal{D}_s$, in the supervised phase, each of the classifiers $\mathcal{C}_1$ and $\mathcal{C}_2$ is trained independently. Formally, given a tweet-label pair $(\tau^i_t, l^i_t)\in \mathcal{D}_s$, a forward pass on the models computes the predicted class probabilities $y^i_1=\mathcal{C}_1(\tau^i_t|\theta_1)$ and $y^i_2=\mathcal{C}_2(\tau^i_t|\theta_2)$. Then two independent gradient-descent based optimizers minimize the following loss functions:
\begin{equation*}
\small
    \begin{split}
        \mathcal{J}_1^s = -\omega_s(l^i_t)\sum_{j=1}^{|\mathcal{L}|} \mu_j\log \nu_{1,j};\ 
        \mathcal{J}_2^s = -\omega_s(l^i_t)\sum_{j=1}^{|\mathcal{L}|} \mu_j\log \nu_{2,j} 
    \end{split}
\end{equation*}
where $\mu_j\in l^i_t$, $\nu_{1,j}\in y^i_1$, $\nu_j\in y^i_2$ are probability of the $j$-th class in the ground-truth, prediction of $\mathcal{C}_1$ and prediction of $\mathcal{C}_2$, respectively. $\omega_s$ is a weighting function to handle class imbalance, defined as:
\begin{equation*}
\small
    \omega_s(l^i_t) = \log \frac{|\mathcal{D}_s|}{|\{l^j_{t'}|l^j_{t'}=l^i_{t}, \tau^j_{t'}\neq \tau^i_t \forall \{\tau^j_{t'}, l^j_{t'}\}\in \mathcal{D}_s\}|+\epsilon}
\end{equation*}
where $\epsilon$ is a very small real value to avoid division by zero.

\subsubsection{Semi-supervised phase}
\label{subsubsec:semi-supervised}
In Figure~\ref{fig:training}, we outline the steps of our proposed semi-supervised phase.
After the purely supervised pass discussed earlier, for any given tweet $\tau^i_t\in \mathcal{D}$, we use the two models to generate pseudo-labels for the last tweets from $f(u_i)$, i.e, the users followed by $u_i$:
\begin{equation*}
\small
    \begin{split}
        y^{f(u_i)_j}_1 = \mathcal{C}_1(\tau^{f(u_i)_j}_{t-}|\theta_1);\
        y^{f(u_i)_j}_2 = \mathcal{C}_2(\tau^{f(u_i)_j}_{t-}|\theta_2)
    \end{split}
\end{equation*}
where $f(u_i)_j$ denotes the $j$-th followee of $u_i$, and $\tau^{f(u_i)_j}_{t-}$ denotes the most recent tweet posted by $f(u_i)_j$ before $t$.

Next, we move on to select labels for the unsupervised part from these pseudo-label sets. One approach can be to compute {\em soft} labels from the sets $\{y^{f(u_i)_j}_1\}_{j=1}^{|f(u_i)|}$ and $\{y^{f(u_i)_j}_2\}_{j=1}^{|f(u_i)|}$ using simple methods like finding the mean. However, similar to \cite{DBLP:conf/nips/MukherjeeA20}, we empirically find that such soft labels do not help much. Instead, we compute {\em hard} or one-hot encoded labels $s^i_1$ ({\em resp.} $s^i_2$) using majority voting from $\{y^{f(u_i)_j}_1\}_{j=1}^{|f(u_i)|}$ $\left(\text{\em resp. }\{y^{f(u_i)_j}_2\}_{j=1}^{|f(u_i)|}\right)$ as follows:
\begin{equation*}
\small
    \begin{split}
        I^{f(u_i)}_1 &= \{\operatorname{argmax}(y^{f(u_i)_j}_1)\}_{j=1}^{|f(u_i)|}\\
        C^{f(u_i)}_1 &= \{\operatorname{freq}(k, I^{f(u_i)}_1)\}_{k=0}^{|\mathcal{L}|-1}\\
        s^i_1 &= \operatorname{one-hot}(\operatorname{argmax}(C^{f(u_i)}_1))
    \end{split}
\end{equation*}
where $\operatorname{freq}(x, X)$ returns the frequency of element $x$ in an array $X$, and $\operatorname{one-hot}(k)$ returns one-hot encoded vector of size $|\mathcal{L}|$ with $k$-th element being $1$.

Additionally, we devise a sample weighing function $\omega_u$ to handle class imbalance in the unsupervised labels. Given a mini-batch of inputs $\mathbf{B}\subset \mathcal{D}$ (see Section~\ref{subsec:implementation} for the mini-batch creation policy), let $\mathcal{S}_1=\{s^i_1\forall t^i_1 \in \mathbf{B}\}$. Then 
\begin{equation*}
\small
    \omega^{\mathbf{B}}_{u, 1}(s^i_1) = \log\frac{|\mathbf{B}|}{\operatorname{freq}(s^i_1, \mathcal{S}_1)+\epsilon}
\end{equation*}

Finally, the pseudo-labels generated by $\mathcal{C}_1$ are used to optimize $\mathcal{C}_2$ and vice-versa. The classifiers predict the class probability distributions $y^i_1$ and $y^i_2$ from the input tweet $\tau^i_t$ in the forward passes of $\mathcal{C}_1$ and $\mathcal{C}_2$, respectively. The objective functions to minimize are:
\useshortskip\begin{equation*}
\small
\begin{split}
    \mathcal{J}_1^u &= -\omega_u(s^i_2)\sum_{j=1}^{|\mathcal{L}|} \sigma_{2,j}\log \nu_{1,j} -\omega_s(l^i_t)\sum_{j=1}^{|\mathcal{L}|} \mu_j\log \nu_{1,j}\\
    \mathcal{J}_2^u &= -\omega_u(s^i_1)\sum_{j=1}^{|\mathcal{L}|} \sigma_{1,j}\log \nu_{2,j} -\omega_s(l^i_t)\sum_{j=1}^{|\mathcal{L}|} \mu_j\log \nu_{2,j}
\end{split}
\end{equation*}
where $\sigma_{1,j}\in s^i_1$, $\sigma_{2,j}\in s^i_2$, $\nu_{1,j}\in y^i_1$, $\nu_{2,j}\in y^i_2$ and $\mu_j\in l^i_t$. The righthand components of the losses correspond to the supervised loss calculated from the annotated label $l^i_t$, if available, and $0$ otherwise.

\subsection{Further implementation details of \name}
\label{subsec:implementation}

As discussed in Section~\ref{subsubsec:semi-supervised}, to compute the pseudo-labels for $\tau^i_t$, we need a forward pass of both the models on the last tweets till $t$ posted by $f(u_i)$. However, if there are multiple consecutive tweets from $u_i$ where most of its followees have not posted a new tweet, we tend to perform redundant, costly forward pass computations. To deal with this, we initialize two matrices $\mathbf{Y}_1$ and $\mathbf{Y}_2$ with the predictions of $\mathcal{C}_1$ and $\mathcal{C}_2$ (right after their supervised training phases), respectively, for the first tweets from each user in $\mathcal{U}$. Then, at each semi-supervised pass, both these matrices are updated with the predictions from $\mathcal{C}_1$ and $\mathcal{C}_2$. We maintain a sparse adjacency matrix of the follow-network for $\mathcal{O}(1)$ time access of $\{y^{f(u_i)_j}_1\}_{j=1}^{|f(u_i)|}$ from $\mathbf{Y}_1$ and $\mathbf{Y}_2$. This way, a single epoch of the semi-supervised training phase is done with $\mathcal{O}(|\mathcal{D}|)$ forward passes for each classifier.

For further speed-up of the training procedure, we opt for mini-batch learning. However, to ensure that $\mathbf{Y}_1$ and $\mathbf{Y}_2$ are updated in a correct temporal order, we maintain the following two conditions:
\begin{enumerate}
    \item Given consecutive batches of input tweets $\mathbf{B}_{k}$ and $\mathbf{B}_{k+1}$, for any pair of tweets $\tau^i_{t1}\in \mathbf{B}_{k}$ and $\tau^i_{t2}\in \mathbf{B}_{k+1}$, $t1<t2$.
    \item For any pair of tweets $\tau^i_{t1}$ and $\tau^j_{t2}$ in the same batch, $u_i\notin f(u_j)$ and vice versa.
\end{enumerate}

Hyperparameter details for implementation of \name\ and the classifiers (learning rate, batch size, minimum followee tweets to threshold, dropout probability, etc.) are detailed in Appendix~\ref{appendix:params} in the supplementary material.

\section{Experiments}
\label{sec:experiments}
\subsection{Dataset}
\label{subsec:dataset}

None of the existing on tweet stance classification datasets include explicit user-user follower information necessary for our experiments. Moreover, due to the ever-changing dynamics of Twitter itself, crawling such additional information related to the old collection of tweets often results in missing data due to deleted tweets, suspended accounts, etc. For this reason, we proceed to collect and annotate our own tweet dataset from scratch. We specifically focused on political tweets from two different demographics: US and India, targeting the opinion dynamics of users around major political parties in these two countries. 
Details of dataset collection and annotation are described in Appendix~\ref{sup:dataset} in the supplementary material.

From US-based users, we collected with a total of $\mathbf{59,684}$ tweets from $\mathbf{24,490}$ users, with an average of $2.44$ tweets and $41.79$ followees per user, within the period from \formatdate{1}{1}{2020} to \formatdate{3}{4}{2020}. We refer to this dataset as \textbf{\usadata}. For the Indian counterpart, our crawling period ranges from \formatdate{26}{10}{2019} to \formatdate{2}{3}{2020}, resulting in a total of $\mathbf{176,619}$ tweets from $\mathbf{63,230}$ users, with an average of $2.79$ tweets and $22.93$ followees per user. We refer to this dataset as \textbf{\indiadata}.

For initial supervised training and final evaluation of \name\ as well as the baselines, we proceed to manually annotate a randomly selected subset of the collected data according to the stance they express toward leading political parties. For US-based tweets, we annotate each tweet with one among the following labels: {\bf Pro-Dem}, {\bf Anti-Dem}, {\bf Pro-Rep}, {\bf Anti-Rep}, and, {\bf Other}. In case of India-based tweets, the labels are {\bf Pro-BJP}, {\bf Anti-BJP}, {\bf Pro-INC}, {\bf Anti-INC}, {\bf Pro-AAP}, {\bf Anti-AAP}, and {\bf Other}. We finally end up with $\mathbf{3,822}$ and $\mathbf{4,185}$ annotated tweets from US and India with inter-annotator agreements $0.84$ and $0.78$ Cohen's $\kappa$, respectively. The distribution of annotated samples among different classes is shown in Table \ref{tab:dataset_class_distr}. To investigate the effects of the size of the labeled training data, we train and test \name\ and all the baselines on three different training sets of size $500$, $1000$, and $1500$.

\begin{table}[!t]
\small
\begin{tabular}{ccccc}
\hline
\multicolumn{1}{l|}{Classes}  & \multicolumn{1}{c|}{Train-500} & \multicolumn{1}{c|}{Train-1000} & \multicolumn{1}{c|}{Train-1500} & Test \\ \hline
\multicolumn{5}{c}{\usadata} \\ \hline
\multicolumn{1}{l|}{Pro-Dem}  & \multicolumn{1}{c|}{320}       & \multicolumn{1}{c|}{654}        & \multicolumn{1}{c|}{981}        & 1543 \\
\multicolumn{1}{l|}{Anti-Dem} & \multicolumn{1}{c|}{9}         & \multicolumn{1}{c|}{22}         & \multicolumn{1}{c|}{32}         & 46   \\
\multicolumn{1}{l|}{Pro-Rep}  & \multicolumn{1}{c|}{133}       & \multicolumn{1}{c|}{253}        & \multicolumn{1}{c|}{381}        & 576  \\
\multicolumn{1}{l|}{Anti-Rep} & \multicolumn{1}{c|}{13}        & \multicolumn{1}{c|}{17}         & \multicolumn{1}{c|}{29}         & 46   \\
\multicolumn{1}{l|}{Other}    & \multicolumn{1}{c|}{25}        & \multicolumn{1}{c|}{54}         & \multicolumn{1}{c|}{77}         & 111  \\ \hline
\multicolumn{1}{l|}{Total}    & \multicolumn{1}{c|}{500}       & \multicolumn{1}{c|}{1000}       & \multicolumn{1}{c|}{1500}       & 2322 \\ \hline
\multicolumn{5}{c}{\indiadata}  \\ \hline
\multicolumn{1}{l|}{Pro-BJP}  & \multicolumn{1}{c|}{67}        & \multicolumn{1}{c|}{149}        & \multicolumn{1}{c|}{208}        & 360  \\
\multicolumn{1}{l|}{Anti-BJP} & \multicolumn{1}{c|}{136}       & \multicolumn{1}{c|}{275}        & \multicolumn{1}{c|}{425}        & 680  \\
\multicolumn{1}{l|}{Pro-INC}  & \multicolumn{1}{c|}{24}        & \multicolumn{1}{c|}{60}         & \multicolumn{1}{c|}{83}         & 158  \\
\multicolumn{1}{l|}{Anti-INC} & \multicolumn{1}{c|}{2}         & \multicolumn{1}{c|}{3}          & \multicolumn{1}{c|}{6}          & 15   \\
\multicolumn{1}{l|}{Pro-AAP}  & \multicolumn{1}{c|}{35}        & \multicolumn{1}{c|}{61}         & \multicolumn{1}{c|}{99}         & 142  \\
\multicolumn{1}{l|}{Anti-AAP} & \multicolumn{1}{c|}{52}        & \multicolumn{1}{c|}{101}        & \multicolumn{1}{c|}{163}        & 210  \\
\multicolumn{1}{l|}{Other}    & \multicolumn{1}{c|}{184}       & \multicolumn{1}{c|}{351}        & \multicolumn{1}{c|}{516}        & 1120 \\ \hline
\multicolumn{1}{l|}{Total}    & \multicolumn{1}{c|}{500}       & \multicolumn{1}{c|}{1000}       & \multicolumn{1}{c|}{1500}       & 2685\\
\hline
\end{tabular}
\caption{Class-wise sample distribution in different training and testing splits of the annotated datasets.}\label{tab:dataset_class_distr}
 \vspace{-5mm}
\end{table}




\subsection{Baselines and ablation variants}
\label{subsec:baseline}

We employ multiple supervised as well as semi-supervised methods to compare with \name. We also remove the contextual view difference and distant network supervision in \name\ step-by-step to investigate their relative importance.
Following are the supervised baselines we implement:
\begin{description}[leftmargin=*]
 \item[SVM.] A linear kernel Support Vector Machine framework with different text-based features, proposed by \citet{Stance-SVM}.
 \item[TAN.] A bidirectional-LSTM based framework with target-specific attention mechanism, proposed by \citet{TAN}.
 \item[SiamNet.] A siamese adaptation of LSTMs, following the works of \citet{SiamNet} and \citet{conforti-etal-2020-will}.
 \item[BICE.] As proposed by \citet{BICE}, this model computes the tweet representation conditioned on targets using Bi-LSTMs.
 \item[BERT.] Pretrained BERT\cite{BERT} (base, uncased) followed by a feed-forward layer for stance prediction.
 \item[ConvNet.] A convolutional model (same as $\mathcal{C}_1$ in \name) trained in a {\em purely supervised} manner using the labeled data only.
 \item[BLSTM.] A bidirectional LSTM model (same as $\mathcal{C}_2$) trained in a {\em purely supervised} manner using the labeled data only.
\end{description}







In the following six semi-supervised baseline methods, we keep the base classifiers among the supervised ones:
\begin{description}[leftmargin=*]
 \item[LP-SVM.] A semi supervised approach based on label spreading \citep{label-spreading} using the SVM model as the base classifier.

\item[ST-ConvNet.] A vanilla self-learning based semi-supervised method using ConvNet as the base classifier. This framework selects $k$-best samples (based on predicted class probability) from the unlabeled dataset and augments the labeled data at each iteration.

\item[ST-BLSTM.] Similar to ST-ConvNet, it replaces the base classifier with the BLSTM model.

\item[UST.] A semi-supervised approach, as proposed by \citet{DBLP:conf/nips/MukherjeeA20} which uses a self-training framework along with BERT for text classification, tailored to our defined categories of stance.

\item[GCN-ConvNet] A semi-supervised node-classification approach similar to \citep{gcn} with ConvNet as node feature extractor followed by two graph convolution layers on the follow-network.

\item[GCN-BLSTM] Similar to GCN-ConvNet with a BiLSTM layer instead of ConvNet. 
\end{description}




\textbf{Ablation variants of \name.}
First, we remove the distant network supervision in \name\ (i.e, no follow network information is used) and reduce it to a vanilla co-training framework with no sophisticated label selection strategy; for each unlabeled instance, the label predicted by $\mathcal{C}_1$ is used to train $\mathcal{C}_2$ and vice versa. We denote this strategy as {\bf \name/Net.} and the resulting classifiers as \name/Net.($\mathcal{C}_1$) and \name/Net.($\mathcal{C}_2$).

Next, to remove the context-based view difference, we use a single classifier in \name\ which computes the majority label from followee tweets for itself. We denote this method as {\bf \name/Cont.} and the resulting classifiers as \name/Cont.($\mathcal{C}_1$) and \name/Cont.($\mathcal{C}_1$).

\section{Results and Discussion}
\label{sec:results}

\begin{table}[!t]
\small
    \centering
    \begin{tabular}{l|c|c|c|c|c|c}
    \hline
        \multirow{2}{*}{Model} & \multicolumn{3}{c|}{\usadata} & \multicolumn{3}{c}{\indiadata} \\ \cline{2-7} 
                       &  \begin{tabular}[c]{@{}c@{}}$|\mathcal{D}_s|$\\ =0.5K\end{tabular}  & \begin{tabular}[c]{@{}c@{}}$|\mathcal{D}_s|$\\ =1K\end{tabular}  &  \begin{tabular}[c]{@{}c@{}}$|\mathcal{D}_s|$\\ =1.5K\end{tabular} & \begin{tabular}[c]{@{}c@{}}$|\mathcal{D}_s|$\\ =0.5K\end{tabular}  & \begin{tabular}[c]{@{}c@{}}$|\mathcal{D}_s|$\\ =1K\end{tabular}  & \begin{tabular}[c]{@{}c@{}}$|\mathcal{D}_s|$\\ =1.5K\end{tabular}   \\ 
    \hline
         SiamNet & $0.39$ & $0.43$ & $0.42$ & $0.12$ & $0.14$ & $0.13$\\
         BICE & $0.27$ & $0.30$ & $0.33$ & $0.16$ & $0.17$ & $0.23$\\
         TAN & $0.38$ & $0.46$ & $0.45$ & $0.14$ & $0.14$ & $0.17$ \\
         SVM & $0.37$ & $0.37$ & $0.45$ & $0.13$ & $0.13$ & $0.16$\\
         BERT & $0.39$ & $0.50$ & $0.51$ & $0.17$ & $0.17$ & $0.21$\\
         ConvNet & $ 0.37 $ & $ 0.43 $ & $ 0.45 $ & $ 0.35 $ & $ 0.40 $ & $ 0.41 $\\
         BLSTM & $ 0.35 $ & $ 0.43 $ & $ 0.44 $ & $ 0.31 $ & $ 0.39 $ & $ 0.38 $\\
         \hline
         LS-SVM & $0.39$ & $0.42$ & $0.44$ & $0.18$ & $0.19$ & $0.18$\\
         ST-ConvNet & $0.13$ & $0.15$ & $0.16$ & $0.10$ & $0.11$ & $0.11$\\
         ST-BLSTM & $0.13$ & $0.16$ & $0.19$ & $0.09$ & $0.12$ & $0.11$\\
         UST & $0.35$ & $0.42$ & $0.41$ & $0.12$ & $0.16$ & $0.16$\\
         GCN-ConvNet & $0.41$ & $0.45$ & $0.47$ & $0.33$ & $0.35$ & $0.40$\\
         GCN-BLSTM & $0.39$ & $0.42$ & $0.46$ & $0.36$ & $0.41$ & $0.42$\\
         \hline
         \name/Net.($\mathcal{C}_1$) & $ 0.32 $ & $ 0.41 $ & $ 0.42 $ & $ 0.10 $ & $ 0.12 $ & $ 0.15 $\\
         \name/Net.($\mathcal{C}_2$) & $ 0.36 $ & $ 0.46 $ & $ 0.46 $ & $ 0.28 $ & $ 0.31 $ & $ 0.37 $ \\
         \name/Cont.($\mathcal{C}_1$) & $ 0.41 $ & $ 0.47 $ & $ 0.49 $ & $ 0.36 $ & $ 0.41 $ & $ 0.43 $\\
         \name/Cont.($\mathcal{C}_2$) & $ 0.47 $ & $ 0.51 $ & $ 0.53 $ & $ 0.38 $ & $ 0.44 $ & $ 0.45 $ \\
         \hline
         \name ($\mathcal{C}_1$) & $\mathbf{0.46}$ & $\mathbf{0.47}$ & $\mathbf{0.49}$ & $\mathbf{0.37}$ & $\mathbf{0.42}$ & $\mathbf{0.45}$\\
         \name ($\mathcal{C}_2$) & $\mathbf{0.49}$ & $\mathbf{0.53}$ & $\mathbf{0.55}$ & $\mathbf{0.42}$ & $\mathbf{0.45}$ & $\mathbf{0.47}$\\
    \hline
    \end{tabular}
    \caption{F1 scores of all models with different sizes of labeled training data on \usadata\ and \indiadata.}
    \vspace{-5mm}
    \label{tab:overall_result}
\end{table}


To evaluate the performance of \name\ along with the baseline models, we use macro-averaged F1 scores. \name\ trains two separate classifiers $\mathcal{C}_1$ and $\mathcal{C}_2$  jointly; however, they are evaluated as independent models on the test sets. We denote them as \name($\mathcal{C}_1$) and \name($\mathcal{C}_2$), respectively.

\subsection{Overall performance}
\label{subsec:overall_performance}

Table \ref{tab:overall_result} presents the performance of \name\ along with the baseline classifiers for three different training splits on the two datasets. 
Clearly, \name\ induces superior classification power in both the classifiers compared to their purely supervised counterparts. The convolutional classifier achieves $0.09$, $0.04$, and $0.04$ points gain in macro-F1 over its supervised version with $500$, $1000$, and $1500$ labelled data, respectively on \usadata. On \indiadata, these gains are $0.02$, $0.02$, and $0.04$, respectively. The performance gains for the bi-LSTM based classifier with \name\ are even more remarkable: $0.14$, $0.10$, and $0.11$
on the \usadata\ while $0.11$, $0.06$, and $0.09$ on \indiadata\ on three different splits respectively. One may relate this disparity in performance gain with the intuitive fact that, bi-LSTMs can capture more complex, long-distance dependencies within the elements of the input text compared to convolutional operations; while small supervised training data hinders this superior representation power to be exploited, with the large volume of data utilized by \name\ empowers the bi-LSTM model to its full potential, thereby accounting for the sharper gain.

Among the external baselines, BERT and BICE emerge as the best-performing ones in US and India datasets, respectively. However, all of these models perform poorly for India-based tweets compared to the US counterpart. A prominent reason behind this difference is the variation in the linguistic quality of tweets from these two demographics. While US-based tweets are purely in English, Indian tweets often contain code-mixed tokens and noisy language. The larger size of the label set in the Indian case is also responsible for the confusion in classification. 

Not surprisingly, vanilla self-training with each of the classifiers (ST-ConvNet and ST-BLSTM) affects the predictive power severely, due to the absence of both contextual as well as text-network view differences. Moreover, these two models do not use any sophisticated label selection methods. With UST, the later problem is resolved with an uncertainty-aware selection strategy. But again, in the absence of network-guided view exploited by \name, UST does not provide any comparable performance. As UST relies purely on textual signals for learning, the performance gap is more evident on \indiadata\ due to a higher degree of noise.

GCN-ConvNet and GCN-BLSTM are the two baselines that, apart from \name, relies on network information. Each of these models show comparative gains with respect to their base counterparts (ConvNet and BLSTM) in most of the cases. However, the models need the network input both at the time of training and testing. This incurs large computation cost at the time of prediction. The classifiers trained using \name\ requires the network-level information only at the time of training to acquire the knowledge about data-distribution from the unlabeled samples. So GCN-based approaches have a relatively higher chance to bias the prediction towards the majority stance of the neighboring nodes. 

\begin{figure}
    \centering
    \includegraphics[width=0.9\columnwidth]{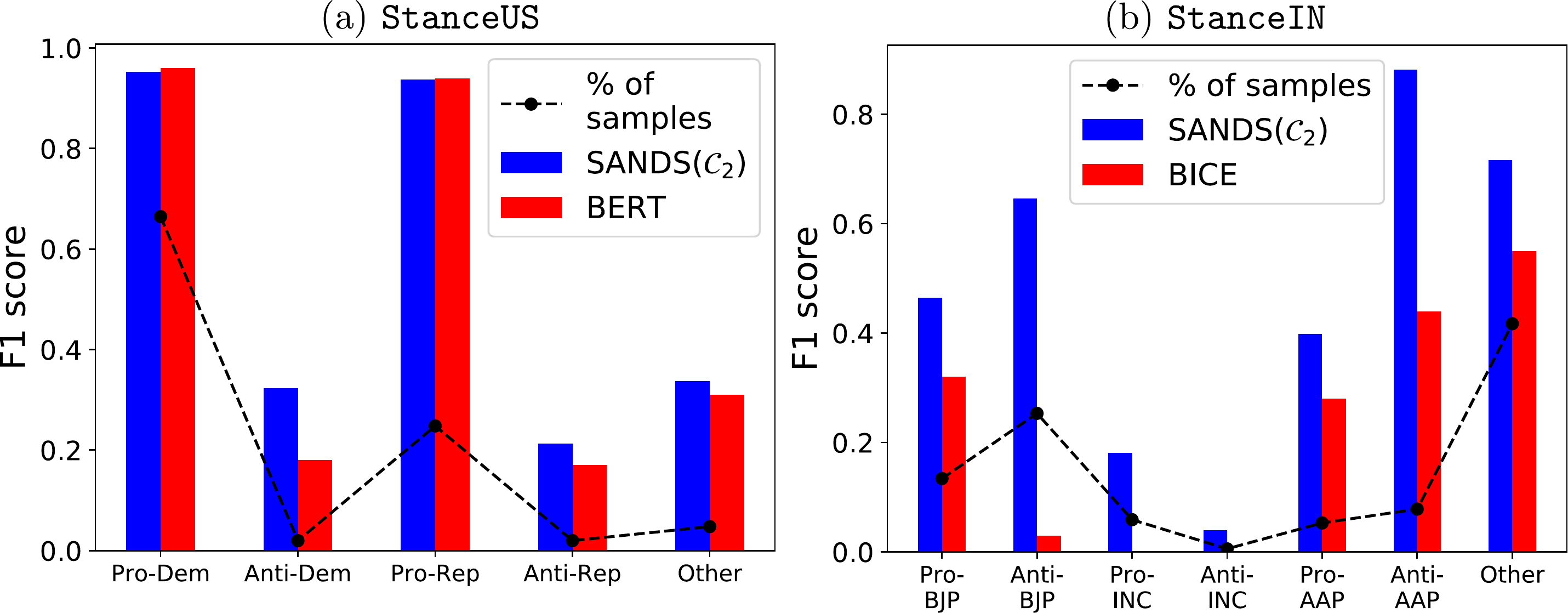}
    \caption{Comparison of class-wise F1 scores between \name($\mathcal{C}_2$) and best supervised baselines -- (a) BERT in US data and (b) BICE in India data.}
    \vspace{-2mm}
    \label{fig:class-wise-baseline}
\end{figure}

\subsection{Handling class imbalance}
\label{subsec:label_bias}

Most real-world datasets suffer from imbalanced ratios of labels, and classifiers that are trained on such data tend to be biased towards dominant classes. This scenario remains the same in our dataset as well, evident from the sample distributions as presented in Table \ref{tab:dataset_class_distr}. 

We start with investigating the effects of class imbalance on \name\ in comparison with the best performing external baselines on each dataset. In Figure \ref{fig:class-wise-baseline}(a), we plot the class-wise F1 scores for \name($\mathcal{C}_2$) vs. BERT, on \usadata\ trained with $1500$ labelled instances. While for the dominant classes like {\bf Pro-Dem} or {\bf Pro-Rep}, both the models perform comparably, in case of minority classes \name($\mathcal{C}_2$) outperforms BERT by a notable margin. Similar is the case on \indiadata, shown in Figure \ref{fig:class-wise-baseline}(b), where the best performing external baseline is BICE. For rarely represented labels like {\bf Pro-INC} or {\bf Anti-INC}, BICE did not predict even a single sample on the test data. 

\begin{figure}
    \centering
    \includegraphics[width=0.9\columnwidth]{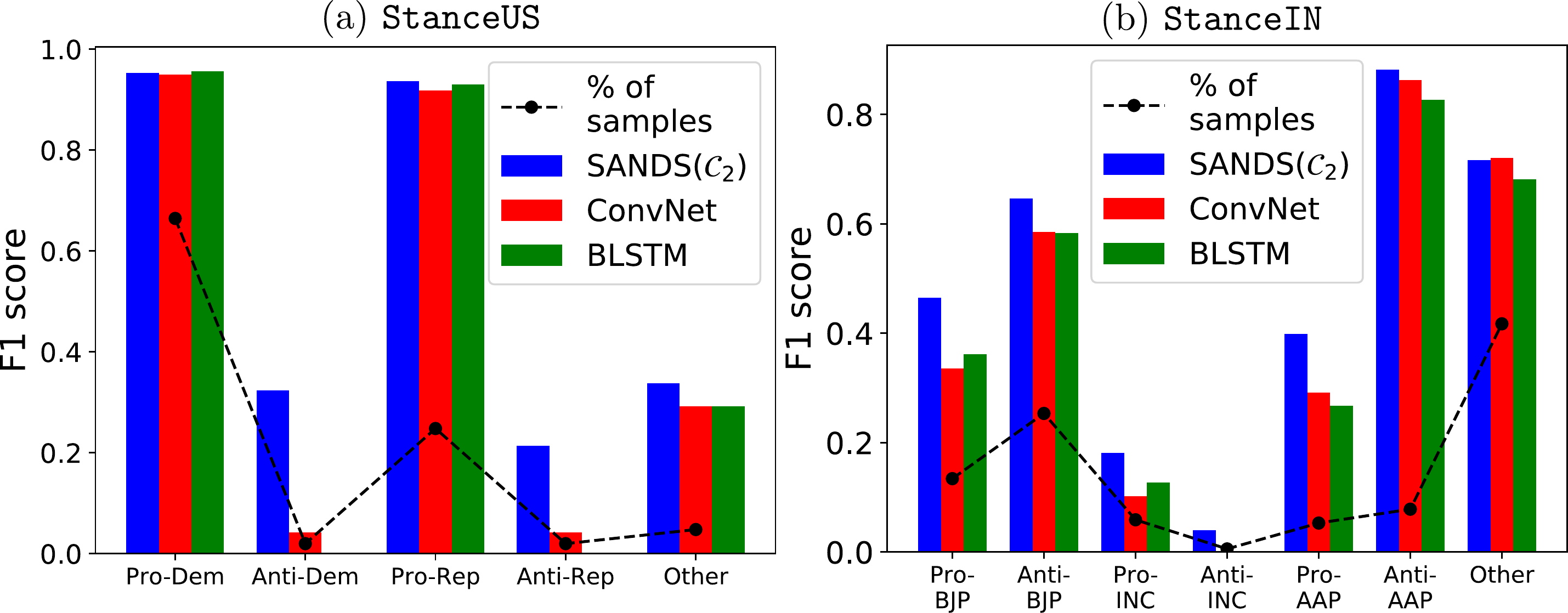}
    \vspace{-4mm}
    \caption{Comparison of class-wise F1 scores of \name($\mathcal{C}_2$) with the supervised counterparts, ConvNet and BLSTM on (a) \usadata\ and (b) \indiadata. All of these frameworks use $1500$ labelled data instances for training. \name\ provides better prediction performance on samples from minority classes compared to the supervised ConvNet and BLSTM.}
    \vspace{-2mm}
    \label{fig:internal_classwise}
\end{figure}

While Figure \ref{fig:class-wise-baseline} provides evidence for the superior performance of \name\ compared to the external baselines for handling biased label distribution, one may suspect whether this betterment is intrinsic to the training methodology of \name, or the classifiers $\mathcal{C}_1$ and $\mathcal{C}_2$ themselves handle the biased samples. To seek further insights, we compare \name($\mathcal{C}_2$) with the purely supervised versions of the classifiers, ConvNet and BLSTM, for class-wise performance. In Figure \ref{fig:internal_classwise}, we can observe that both classifiers suffer heavily due to label imbalance. The supervised BLSTM model can not predict any sample from the two minority classes in \usadata, while ConvNet performs very poorly. In both cases, \name\ provides substantial betterment. Similarly on \indiadata, where both the supervised classifiers could not predict any samples with the {\bf Anti-INC} label while \name$(\mathcal{C}_2$), (although poorly) provides some.

All of the supervised classifiers (external as well as ConvNet and BLSTM) take the class imbalance manifested within training samples into account with some sample weighting strategy (similar to $\omega_s$ in Section \ref{subsubsec:supervised_pretrain}). Instead of this, these classifiers perform miserably on minority labels compared to \name. The vast majority of unlabelled data used in \name\ provides $\mathcal{C}_1$ and $\mathcal{C}_2$ with stronger feature signals for such labels that are absent in the small amount of labeled training data. Moreover, the loss weighing parameter $\omega_u$ (see Section \ref{subsubsec:semi-supervised}) devised in \name\ forces the models to pay attention to the samples with under-represented stance labels.
\begin{figure}[!t]
    \centering
    \includegraphics[width=\columnwidth]{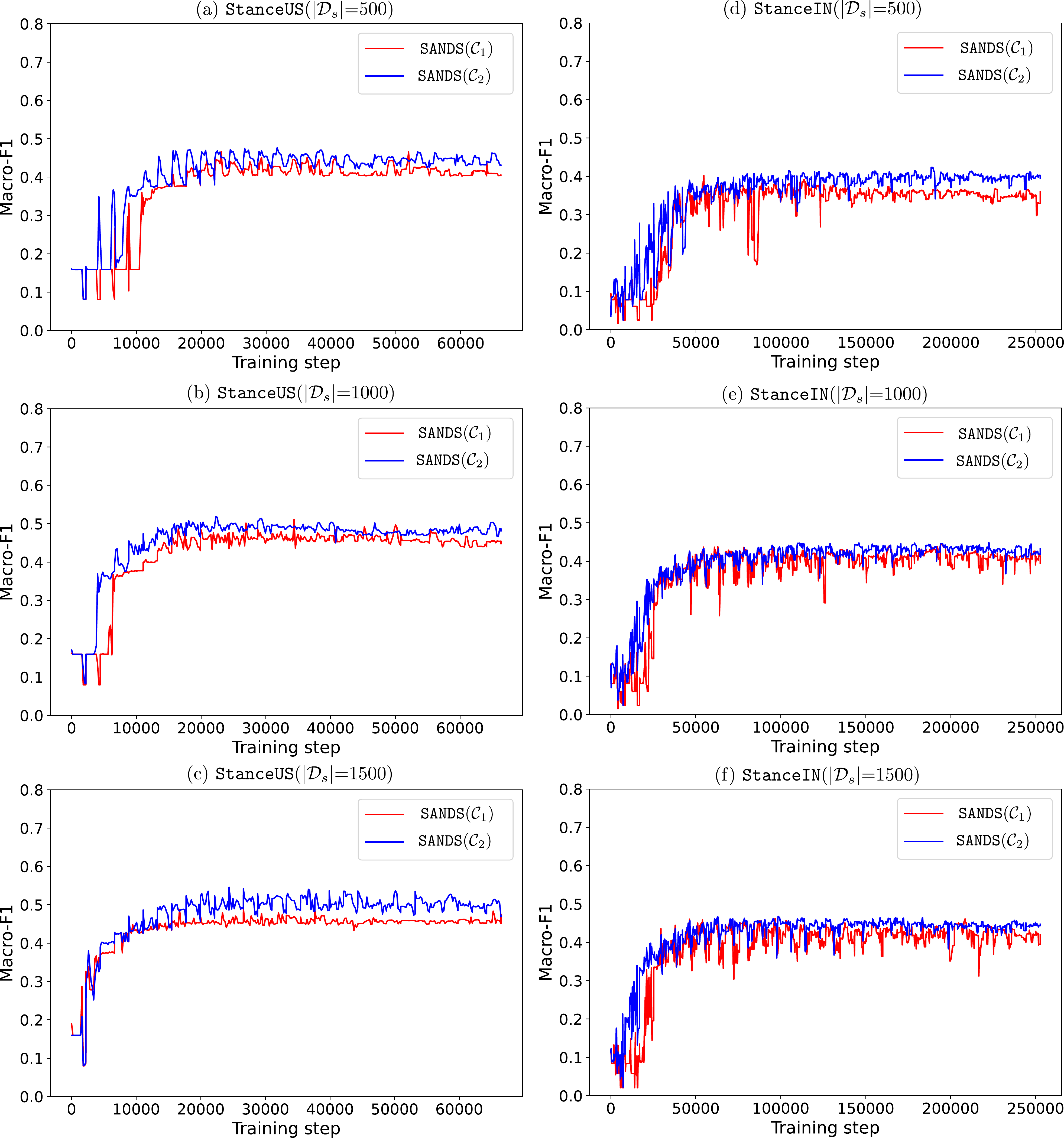}
    \vspace{-3mm}
    \caption{Macro-F1 scores of \name($\mathcal{C}_1$) and \name($\mathcal{C}_2$) on test set as training progress on different training splits for \usadata\ and \indiadata. In each case, \name\ provides stable convergence.}
    \label{fig:training_stability}
    \vspace{-2mm}
\end{figure}

\subsection{Importance of network and feature context}
\label{subsec:ablation-results}

The relative importance of distant supervision from network dynamics and synthetic difference of contextual views imposed by the classifier designs can be interpreted from the performances of \name/Net. and \name/Cont. classifiers in Table~\ref{tab:overall_result}. For both datasets, the performance drops for each of the ablation variants from the original setting are evident. However, without distant network supervision, the classifiers suffer more severely: $0.07$ ($0.09$) points drop in macro-F1 score $\mathcal{C}_1$ ($\mathcal{C}_2$) compared to $0.0$ ($0.02$) points drop in the absence of contextual view difference on \usadata. 

The absence of network supervision affects the performance more drastically in the case of \indiadata. The predictions of the convolutional classifier ($\mathcal{C}_1$) are essentially random. While both the ablation variants show unstable behavior while training progresses, the fluctuations in \name/Net. models on \indiadata\ are much significant. This difference of effects of network supervision over US and India-based tweets further strengthens our previous observation --- network dynamics play a more important role for highly noisy tweets in the Indian context.

\subsection{Stability and convergence of learning}
\label{subsec:stability}

Over and above performance gains, 
a training protocol must provide stable, convergent learning of the underlying model(s). To investigate \name\ for this property, we continue training for a sufficient number of optimization steps and track the performance of each of the classifiers $\mathcal{C}_1$ and $\mathcal{C}_2$ on the test set. In Figure \ref{fig:training_stability}, we plot the resulting progress for each of the datasets with different amount of available labeled training data. We can observe that, in each case, both the models reach a stable convergence overcoming the periodic fluctuations at the initial phases. However, these fluctuations are not classifier-specific and vary with datasets as well as the size of labeled data. For example, on \indiadata, $\mathcal{C}_1$ shows more fluctuations in performance even at the later stages, compared to $\mathcal{C}_2$. However, in the US counterpart, this behavior reverses with a more stable convergence of $\mathcal{C}_1$ than $\mathcal{C}_2$. In both the datasets, with more labeled data, the classifiers reach stability faster. Such dependence is trivial, as with smaller labeled data the classifiers search around the parameter space in a more unguided manner.

\subsection{Choice of classifier pairs}
\label{subsec:classifier_pair}

\begin{table}[!t]
\small
\centering
\begin{tabular}{l|c|c|c|c}
\hline
\multirow{2}{*}{\begin{tabular}[c]{@{}l@{}}Classifier pair\\ ($\mathcal{C}_1$, $\mathcal{C}_2$)\end{tabular}} & \multicolumn{2}{c|}{\usadata} & \multicolumn{2}{c}{\indiadata} \\ \cline{2-5} 
                                                                                    & m-F1 $\mathcal{C}_1$     & m-F1 $\mathcal{C}_2$    & m-F1 $\mathcal{C}_1$      & m-F1 $\mathcal{C}_2$     \\ \hline
Conv, bi-LSTM                                                                       & 0.49        & 0.55       & 0.45         & 0.47        \\
Conv, Conv                                                                          & 0.48        & 0.48       & 0.43         & 0.43        \\
bi-LSTM, bi-LSTM                                                                    & 0.52        & 0.52       & 0.45         & 0.45        \\
Conv, BERT                                                                          & 0.50        & 0.52       & 0.44         & 0.45        \\
BERT, bi-LSTM                                                                       & 0.53        & 0.53       & 0.42         & 0.45        \\ \hline
\end{tabular}
 \caption{Macro-F1 scores of different classifier-pairs with \name.
The size of labelled dataset used for all these pairs is $1500$.}
\vspace{-5mm}
\label{tab:classifier_pair}
\end{table}

We experiment with multiple different choices of classifier designs for $\mathcal{C}_1$ and $\mathcal{C}_2$ with \name\ and summarize the results in Table \ref{tab:classifier_pair}. One can observe the differences in the performance of the same classifier when paired with two different ones; e.g., a convolutional classifier performs better when paired with BERT instead of bi-LSTM. However, neither of these two classifiers could outperform bi-LSTM when paired with a convolutional classifier. 
When the same design is chosen for both classifiers, we remove the view-difference enforced by feature signal encoding from local and global contexts, and the settings become essentially similar to the ablation variant \name/Cont., showing a very similar performance to \name/Cont.($\mathcal{C}_1$) and \name/Cont.($\mathcal{C}_2$).

\begin{table}[!t]
\small
\centering
\begin{tabular}{c|c|c|c|c}
\hline
\multirow{2}{*}{\begin{tabular}[c]{@{}l@{}}\% unlabeled\\ data used\end{tabular}} & \multicolumn{2}{c|}{\usadata} & \multicolumn{2}{c}{\indiadata} \\ \cline{2-5} 
                                                                                    & m-F1 $\mathcal{C}_1$     & m-F1 $\mathcal{C}_2$    & m-F1 $\mathcal{C}_1$      & m-F1 $\mathcal{C}_2$     \\ \hline
$100$                                                                       & 0.49        & 0.55       & 0.45         & 0.47        \\
$80$                                                                          & 0.49        & 0.54       & 0.44         & 0.46        \\
$50$                                                                    & 
0.46        & 0.51       & 0.42         & 0.43        \\
$30$                                                                          & 0.45        & 0.48       & 0.42         & 0.41        \\
$10$                                                                       & 0.45        & 0.45       & 0.41         & 0.39        \\ \hline
\end{tabular}
\caption{Macro-F1 scores of \name\ with different amount of unlabeled data used in semi-supervised phase. The size of labeled dataset used is $1500$.}
\label{tab:vary_data_size}
\vspace{-4mm}
\end{table}
\subsection{Variation in the size of unlabeled data}
\label{subsec:data_variation}

Finally, we seek to explore how the amount of unlabeled data used in \name\ effects the performance of each of the classifiers. We experiment with $80\%$, $50\%$, $30\%$, and $10$\% of the original amount of unlabeled data for both the datasets and present the results in Table~\ref{tab:vary_data_size}. In both cases, the effect of data reduction manifests more sharply over the performance of $\mathcal{C}_2$, i.e., the bi-LSTM classifier. This is at par with the fact that $\mathcal{C}_2$ received a sharper gain in predictive power from supervised to semi-supervised setting (see Section~\ref{subsec:overall_performance}). Also, the dip in performance with decreasing amount of unlabeled data is more evident on \indiadata\ compared to its US counterpart. 

\section{Conclusion}
\label{sec:conclude}
We presented \name, a novel stance classification framework that uses distant supervision by social network properties to exploit large-scale unlabeled data along with very few manually labeled data, to predict political stance expressed over tweets.
We experimentally validated the superiority of our proposed method in comparison to several supervised as well as semi-supervised classification frameworks on two tweet stance datasets we collected and annotated from US and India-based users. Particularly in settings with small annotated data with noisy textual representations and highly imbalanced class-wise sample distribution, \name\ improved upon other models by substantial margins. With different ablation experiments, our work provides significant insights into the complex interplay of text-based and network-propagated signals for stance classification.

\bibliographystyle{ACM-Reference-Format}
\bibliography{ref1}

\newpage
\thispagestyle{plain}
\makeatletter
\twocolumn[\centering \@titlefont \ztitle \\
\vspace{.6ex}
\LARGE (Appendix / Supplementary Material) \par \bigskip
\makeatother ]
\appendix
\begin{figure}[!t]
    \centering
    \includegraphics[width=\columnwidth]{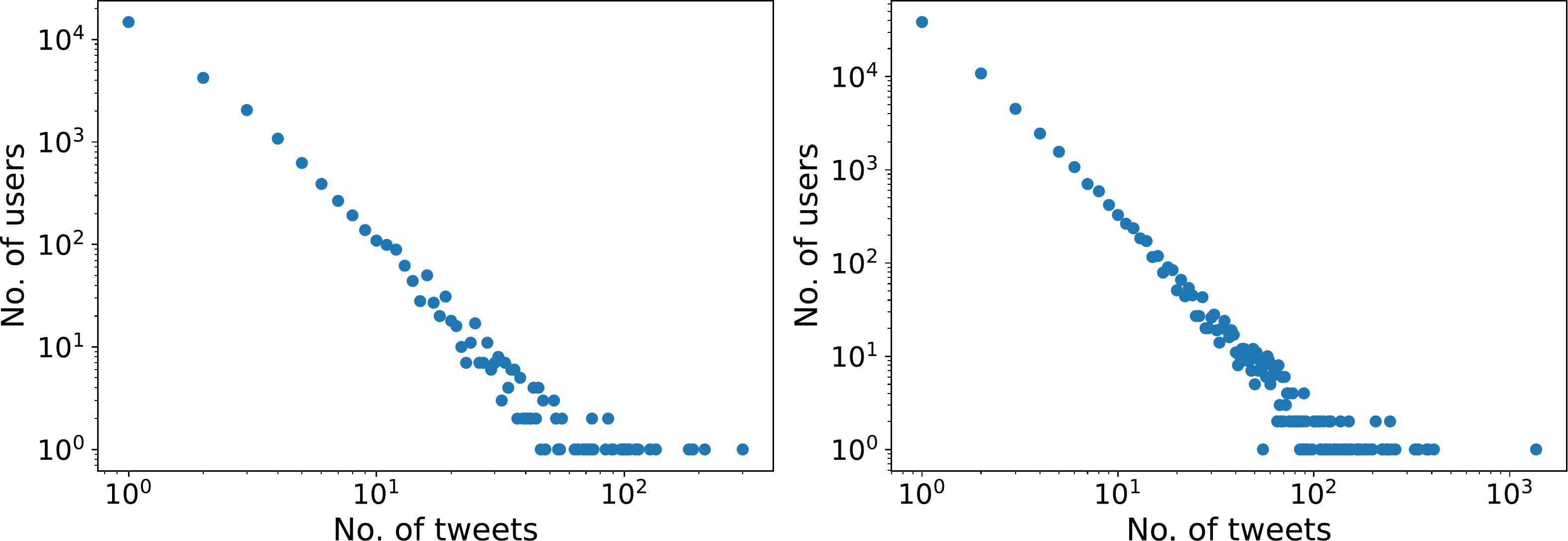}
    \caption{Log-log distribution of tweets posted by users in \usadata\ (left) and \indiadata\ (right).}
    \label{fig:loglog-data}
\end{figure}
\begin{figure*}[h]
    \centering
    \includegraphics[width=\textwidth]{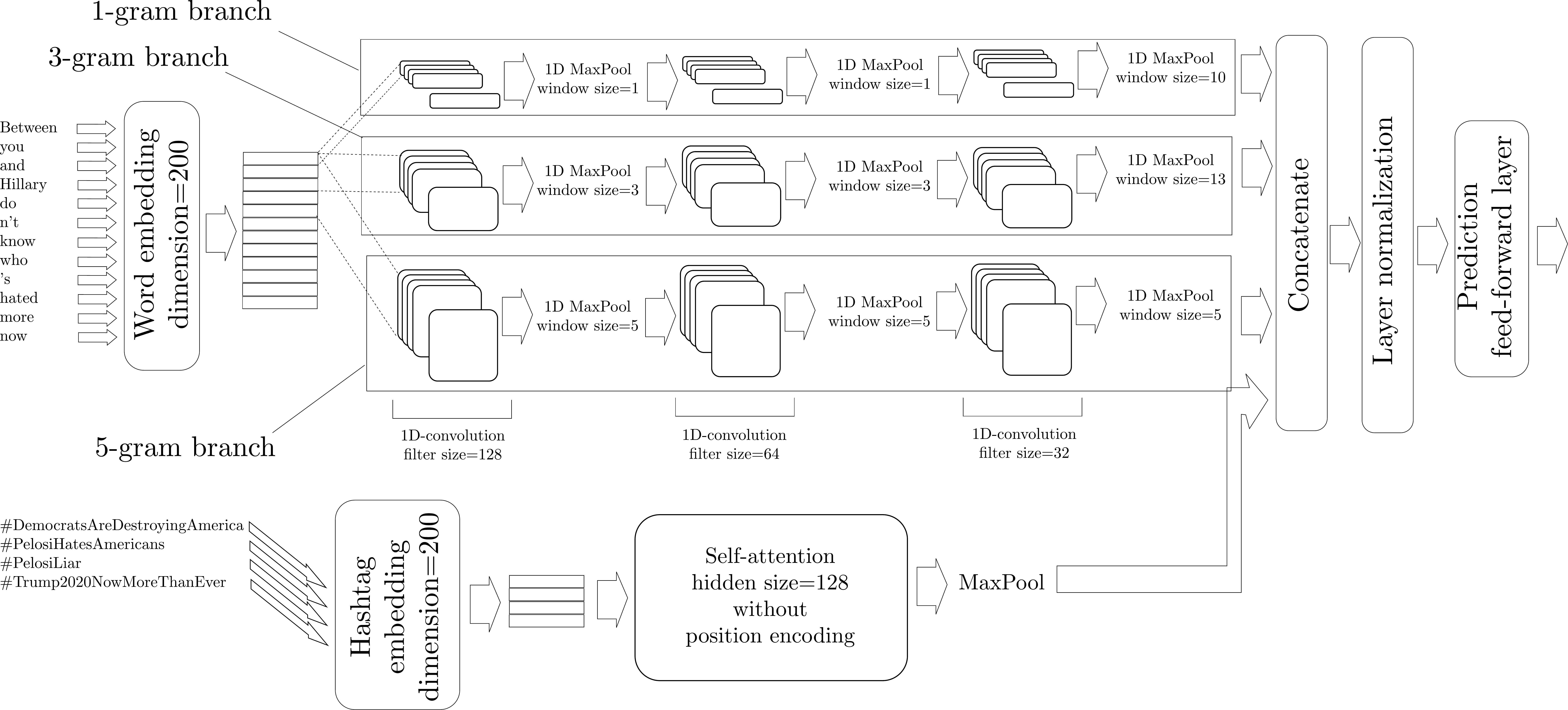}
    \caption{Fully expanded view of the convolutional model}
    \label{fig:my_label}
\end{figure*}
\section{Dataset preparation}
\label{sup:dataset}

To collect US political stances, we first identified a set of $100$ users who expressed support towards the Democratic (Dem) or the Republican (Rep) Parties explicitly on their profile information ($50$ users each). We also kept track of daily trending hashtags (political as well as non-political) and collected tweets that mentioned them. We then created the full social network comprised of the said $100$ users and the users that posted using the trending hashtags. We continued this procedure from \formatdate{1}{1}{2020} to \formatdate{3}{4}{2020}. We also collected tweets posted by a user in our collection within this time-frame which was not collected while searching for trending hashtags to ensure continuity of tweeting activity. Finally, we end up with a total of $\mathbf{59,684}$ tweets from $\mathbf{24,490}$ users, with an average of $2.44$ tweets and $41.79$ followees per user. We refer to this dataset as \textbf{\usadata}.

For the Indian counterpart, we followed the same procedure. However, we focused mostly on the capital city area of New Delhi to ensure most tweets to be written in English. The targeted political bodies for this demographic area were Bharatiya Janta Party (BJP), Indian National Congress (INC), and Aam Admi Party (AAP). Our crawling period ranges from \formatdate{26}{10}{2019} to \formatdate{2}{3}{2020}, resulting in a total of $\mathbf{176,619}$ tweets from $\mathbf{63,230}$ users, with an average of $2.79$ tweets and $22.93$ followees per user. We refer to this dataset as \textbf{\indiadata}.
Figure~\ref{fig:loglog-data} shows the distribution of the number of tweets posted by users, evidently following power-law.

For initial supervised training and final evaluation of \name\ as well as the baselines, we proceed to manually annotate a randomly selected subset of the collected data according to the stance they express. For US-based tweets, we annotate each tweet with one among the following labels: {\bf Pro-Dem}, {\bf Anti-Dem}, {\bf Pro-Rep}, {\bf Anti-Rep}, and, {\bf Other}. In case of India-based tweets, the labels are {\bf Pro-BJP}, {\bf Anti-BJP}, {\bf Pro-INC}, {\bf Anti-INC}, {\bf Pro-AAP}, {\bf Anti-AAP}, and {\bf Other}. We employ three expert annotators for this task (their ages range between 25--35 years). All of them are from a linguistics background and familiar with political events and entities in both demographics.  In case of disagreement among the annotators, we select the majority label (if two of them agree), else discard it altogether. After a $5$-week long annotation process, we finally end up with $\mathbf{3,822}$ and $\mathbf{4,185}$ annotated tweets from US and India with inter-annotator agreements $0.84$ and $0.78$ Cohen's $\kappa$, respectively. The distribution of annotated samples among different classes is shown in Table \ref{tab:dataset_class_distr}.

To investigate the effects of the size of the labeled training data, we train and test \name\ and all the baselines on three different training sets of size $500$, $1000$, and $1500$.

\section{Tweet preprocessing}
\label{appendix:preprocess}

For both \usadata\ and \indiadata, we use the same strategy for cleaning the tweets. The text of each tweet is converted to lowercase. The URLs in the tweet are removed. All URLs are first found using regular expression search and then, replaced by a space. We further, created a list of hashtags used in the tweet and removed any hashtags or mentions that were present in the tweet. Finally, we remove any punctuation or numbers present in the tweet.

In order to create the vocabulary of words, we begin by tokenizing the tweets and creating a set of all the tokens. From this set, all non-ASCII terms are removed, along with the terms that occur only once. For obtaining the vocabulary of hashtags, from the set of all hashtags, the hashtags that occur less than or equal to five times are removed to exclude any rare hashtags. After this, we obtained a vocabulary of size $16166$ and hashtag vocabulary of size $2134$ for \usadata. The maximum length for the tweets (number of terms in the cleaned tweet) in the \usadata\ is $59$ and the maximum hashtag length is $30$. For \indiadata, the word vocabulary size is $31939$ and the size of hashtag vocabulary obtained is $3480$. For the cleaned tweets in \indiadata, the maximum tweet length is found to be $61$ and the maximum length for the hashtags in a given tweet is $28$.

\section{Parameter details}
\label{appendix:params}
In both the models (convolutional and bi-LSTM) in our setting, the word embedding layers are initialized with \href{http://nlp.stanford.edu/data/glove.twitter.27B.zip}{pretrained GloVe}
word vectors of dimension $200$. The hashtag embedding layers were initialized randomly with size $128$. We keep all these embeddings trainable throughout the training process. 

For the self-attention based module to represent hashtags (see Section~\ref{subsubsec:hashtag}), we set the dimensions of the query, key, and value vectors to $128$, which then trivially becomes the dimensionality of $\mathbf{Z}^H$ as well.

In the three successive convolution-maxpool blocks (see Section~\ref{subsubsec:conv_model}), we keep the filter sizes as $128$, $64$, and $32$, respectively, in all the branches of different window sizes. In the bi-LSTM model, we set the hidden size of the LSTM cell to $200$ ($100$ in each direction and concatenated). 

Each training iteration in our method consists of a supervised pass followed by a semi-supervised one (as described in Section~\ref{subsec:training}). In the supervised pass, we set the dropout probability $p_{\text{dropout}}=0.1$ (see Section~\ref{subsec:models}) with a batch size of $128$. Due to the constraints described in Section~\ref{subsec:implementation}, the mini-batches in the semi-supervised pass are of variable sizes and we set the upper bound on the size to be $512$. We set $p_{\text{dropout}}=0.3$ in this pass. We use Adam optimizer with a learning rate of $10^{-4}$ for both the models.

As follower-network information for Twitter users is often kept private due to profile settings, sometimes it is not possible to collect the information about all the followees of a given user. To ensure that the pseudo-labels computed from the tweets posted by the followees $f(u_i)$ (see Section~\ref{subsubsec:semi-supervised}) do not get biased due to the small size of followees, we use an empirically set $min\_degree$ threshold to ignore pseudo-labels for users having followees less than that value. We set $min\_degree=15$ for \usadata\ and $20$ for \indiadata.
\end{document}